# LASER-INDUCED BREAKDOWN SPECTROSCOPY: A VERSATILE TECHNIQUE OF ELEMENTAL ANALYSIS AND ITS APPLICATIONS

V. N. Rai

Raja Ramanna Centre for Advanced Technology

Indore – 452 013 (India)

Email: vnrai@rrcat.gov.in

## ABSTRACT

This paper reviews the state of art technology of laser induced breakdown spectroscopy (LIBS). Research on LIBS is gaining momentum in the field of instrumentation and data analysis technique due to its wide application in various field particularly in environmental monitoring and in industry. The main focus is on its miniaturization for field application and on increasing its sensitivity. The sensitivity of LIBS has been increased by confining the laser produced plasma using external magnetic field as well as using two successive laser pulse excitation of plasma. LIBS has capability for simultaneous multi element determination, localized microanalysis, surface analysis and has been used successfully for determination and identification of hazardous explosive and biological samples. Experimental findings of LIBS study in different applications have been discussed.

## 1.  INTRODUCTION

Laser-induced breakdown spectroscopy (LIBS) is a laser diagnostics, where a laser beam focused onto a material generates transient high-density plasma as the laser intensity exceeds the breakdown threshold of the material (~1-10 $MW/cm^2$). The UV and visible emission from the plasma can be spectrally resolved and recorded for qualitative and quantitative analysis of the sample. LIBS was first used for the determination of elemental composition of materials in the form of gases, liquids and solids during 1960's[1-2]. Research on LIBS continued to grow and reached a peak around 1970 and field-portable instruments capable of in-situ and real time analysis of samples have been developed in recent years with the availability of reliable, smaller and less costly laser systems along with sensitive optical detectors, such as the intensified charge-coupled device (ICCD). Several review articles have been published on this topic[3-14]. In the LIBS, the wavelength of light provides information on the identity of elements composing the sample, while the intensities reflect the relative abundance of these components within the samples. Once the LIBS system is coupled to a broadband detector such as Echelle spectrometer with intensified CCD detector or recent multi spectrometer system, it provides a broad band spectrum (200- 900 nm) of the elements present in the sample in real time.

LIBS has many advantages as an analytical technique. There is no need of sample preparation, which avoids further contamination of the material to be analyzed[10-11]. The analysis process is fast and can be used for both non-conducting and conducting samples, regardless of their physical states, i.e. aerosols, gases, liquids or solids. LIBS is also applicable to the analysis of extremely hard materials that are difficult to digest or dissolve, such as ceramics and semi or super-conductors. Its capability for simultaneous multi-element determination, localized microanalysis, and surface analysis are also of great importance and it has been used successfully in hazardous and difficult environmental conditions to study remotely located samples for online and real time information about their spectra. LIBS has been found useful in elemental process monitoring and in field portable analyzers for in situ trace metal analysis of real samples where accuracy and precision are not the main requirement[11].

More recently it has been demonstrated that LIBS can be applicable in several military and homeland security applications[15-18]. Nowadays these applications has important role because the threat of chemical and biological terrorist acts have increased the demand for accurate and rapid determination of chemical and biological warfare (CBW) agent species[19]. The identification of these chemical and biological agents is difficult because these agents exist in various forms such as liquid, vapor and fine powders. These agents may be present in low concentration relative to background interferant such as dust, soil or painted surfaces as well as in air. Handling these agents are difficult so LIBS has been found as a suitable technique for detection and identification of chemical and biological warfare (CBW) agents. The application of LIBS as a field portable detector for CBW agents requires the development and optimization of statistical methods for rapidly analyzing complex spectra obtained in the field. Statistical methods have been used to interpret spectroscopic data in many other applications. Recently some groups of researcher have demonstrated the feasibility of biomaterial identification using LIBS coupled with chemometric methods.

This article presents briefly the technical aspect of LIBS along with its various applications in different fields, which will be beneficial in understanding the environmental conditions of the planet Earth.

## 2. SCOPE OF LIBS

The focal theme of 94th Indian Science Congress is Planet Earth. We all are aware that our planet earth has a beautiful environment, which helps in maintaining the life on it. Particularly there are oceans full of life, and lands to grow food on. Polar ice caps are reserving water. Plants and trees provide oxygen, where as atmosphere collects warmth from the Sun. Swirling clouds recycle moisture and provide fresh water. But now a days all parts of the earths environment such as land, water, atmosphere, weather and plants has problems mainly due to the fast growth of industrialization, which are observed in the form of pollution in soil, water and air as well as in the form of global warming, where as fast deforestation creates loss of oxygen in the atmosphere. Pollution are mainly due to the presence of hazardous high atomic number elements such as Pb, As, Hg, Cr, Cd and Be etc. in solid, liquid and gaseous materials. Hazardous and toxic elements as well as compounds can harm the life in different ways. Due to this an easy technique was required since long, which can give information about the presence of these harmful agents in the earth's environment. LIBS has been identified as a laser diagnostics technique for detection of hazardous and toxic elements in any kind of base material. This technique is equally useful for detecting the composition of materials in alloy industry as well as in finding the trace elements present in different environmental conditions. Recently it has been found useful in detection and identification of compounds of any materials particularly hazardous chemical and biological material. Detection and identification of compounds are difficult than the elemental analysis.

This technique is mainly based on collecting the emission from ionic, atomic and molecular species in a plasma spark. All elements emit in the spectral range of 200-900 nm. This indicates that by using a detector covering the entire spectral range, one can detect and identify elements and molecular species by their relative abundance. The complex compound of explosive consists of a fuel component such as hydrocarbon and an oxidizer such as oxides of nitrogen contained in the same molecular structure. The requirement for detecting and identifying these chemical explosive, one need to identify wide variety of different constituents present in the explosives materials.

Similarly as explosives, hazardous biological and bacteriological materials have become the subject of concern due to the increased terrorist activities. Detection and identification of hazardous biological samples are important not only to avoid terrorist activity but in civilian application also such as bacteriological monitoring in medical sciences as well as in hygienic control in food processing industries. Hazardous biological samples include mainly bacteria, viruses, biotoxins and fungi. Very low concentration of these materials can create large-scale contamination. Biological samples consists of bacterial cells, which is a living structure composed of a genome, cytoplasm and various membranes. It is a biological structure with chemical elements in complex molecules and is organized in a sophisticated manner related to its function as part of living organism such as in metabolism, reproduction, enzymatic activity, respiration etc. Due to its complex structure it is desirable to find some typical atomic markers that are reliable enough to give some information on the nature of living organisms and their potential pathogenic capabilities. The molecules of the samples are completely dissociated during the plasma formation process, where mainly ionic and atomic species are present in the plasma. The characteristics de-excitation emission lines from ionic and atomic species are used for detection in less than 1 ms. These qualitative and quantitative spectrochemical analysis provide information for the identification of biological samples. This is a challenging job because of the complex biological structure and its composition. However with current advances in broadband detectors such as multi spectrometer or Echelle spectrometer, LIBS is capable of detecting a wide variety of toxic and hazardous compounds, which is not possible by any other technique. LIBS has many other applications in different field of research. Many aspects of LIBS along with some of its important applications will be discussed briefly in the following text.

## 2.1   Spectral Lines of Interest

The application of LIBS in detection and identification of hazardous chemical and biological samples (compounds) depends mainly on the choices of characteristics spectral markers. In the case of hazardous chemical explosive main elements present in the spectrum are carbon (C), hydrogen (H), nitrogen (N), and oxygen (O), which provide major peaks. In the case of inorganic explosive such as black powder, spectrum is richer in peaks because of presence of inorganic components particularly K, Ca.

In the case of biological samples, both types of elements inorganic and organic are detected. Inorganic elements include magnesium (Mg), sodium (Na), iron (Fe), Potassium (K) and Calcium (Ca). Presence of these elements can be used as marker. However Na and Ca are not of much interest because of their presence in atmosphere. Detected organic elements include carbon (C), Nitrogen (N), phosphorous (P) and hydrogen (H). Out of these C is important as an organic marker.

## 3. EXPERIMENTAL SETUP FOR LIBS

Various types of experimental set up have been used for LIBS, where different types of collection optics are used for the radiation emitted by the plasma plume. Sometime emission from the plasma is collected in the direction perpendicular to the direction of the incident laser. In addition to the difficulties of alignment and reduced sensitivity, the collected emission exhibits spatial dependence, leading to loss of spectral information about emission from the whole plasma plume. These shortcomings are removed in another arrangement, where focusing lens itself acts as the collecting lens for the plasma emission. In this case collected emission corresponds to the integrated value of radiation from all the spatial locations of the plasma. The typical schematic diagrams of the experimental set-ups[14-15, 21-22] for recording the laser-induced breakdown emission from the solid, liquid and gaseous samples are shown in Figs 1, 2 and 3 respectively.

The LIBS experimental set up for studying solid samples consists of a Q-switched, frequency-doubled Nd: YAG laser (Continuum Surelite III) that delivers energy of ~ 300 mJ at 532 nm in 5-ns pulse. This laser was operated at 10 Hz and was focused on the target with the help of a dichroic mirror and quartz-focusing lens of 20 cm focal length. The combination of dichroic mirror and the same focusing lens (Fig.-1) was used to collect the optical emission from the laser-induced plasma. Two UV grade quartz lenses of focal lengths 100 mm and 50 mm were used to couple the plasma emission to an optical fiber bundle. The fiber bundle consists of 80 single fibers of 0.01 mm core diameter. The rectangular exit end of the optical fiber was coupled to the spectrograph (Model HR 460, Instrument SA, Inc., Edison, NJ) and used as an entrance slit. The spectrograph was equipped with 1200 and 2400-l/mm diffraction gratings of dimension 75 mm × 75 mm. A 1024 × 256 element intensified charge-coupled detector (ICCD) (Princeton Instrument Corporation, Princeton, NJ), with a pixel width of 0.022 mm, was attached to the exit focal plane of the spectrograph and used to detect the dispersed light from the laser-induced plasma. This spectrograph provides limited spectral bandwidth. One has to change the grating position time to time for recording the whole band of spectrum. The detection and identification of hazardous chemical and biological samples require simultaneous recording of broadband spectra. For this purpose one has to use different types of spectrometer, detail of which is provided in the following text. The detector was operated in gated mode with the control of a high voltage pulse generator (PG-10, Princeton Instruments Corporation, Princeton, NJ) and was synchronized to the output of the laser pulse. Data acquisition and analysis were performed using a personal computer. The gate delay time and gate width were adjusted to maximize the signal-to-background (S/B) and signal-to-noise (S/N) ratios. Emission spectra were recorded mainly using 2400-l/mm grating for a better spectral resolution. Around 100 pulses were accumulated to obtain one spectrum and 30 such spectra were recorded for each experimental condition in order to increase the sensitivity of the system and to reduce the standard deviation in the recorded data. LIBS experiments can be performed in air, low-pressure inert gas, and vacuum. Some considerations concerning the construction of vacuum and gas chambers mainly include two aspects: (i) to extend the analytical spectral range to the deep-UV region for elements such as carbon, phosphorous, sulfur, chlorine, bromine, iodine, oxygen, and nitrogen; and (ii) to improve the detectability by inert gas purging. Research on laboratory applications of LIBS are carried out either on a sample kept in vacuum chamber or in open atmosphere, but field applicable LIBS is usually performed in the atmospheric environment.

## 3.1   Broad Band Spectrometers for LIBS

Czerny- Turner spectrographs are usually employed to disperse the emission collected from LIBS plasma, where suitable detectors coupled to it offer the possibility of time resolved measurements. These detection systems are intrinsically limited in resolution as well as in spectral coverage. The multi-elemental detection capability offered by the LIBS technique demands a spectrograph with a wider spectral coverage. In the multi-elemental analysis, sequential measurements of parts of the spectrum of interest are performed, inspecting each time a different sample of the material ablated from the target surface. In principle this procedure limits the LIBS application to homogeneous samples. But most of the samples are inhomogeneous, which is why the spectra vary from shot to shot, as a result of changes in the sample composition as well as due to stochastic fluctuations in the plasma.[23] Particularly detection and identification of hazardous chemical and biological samples also require simultaneous recording of broad band spectra for getting optimum information for analytical purposes. Instruments, which allow simultaneous measurements, are Paschen - Runge spectrometers or the more compact Echelle spectrometers as discussed in a review article by Detalle et al.[24] Echelle spectrometer offers excellent spectral resolving power ($\lambda / \Delta\lambda \geq 10.000$ and more) in combination with a spectral coverage of several hundred nanometers. In combination with intensified charge coupled devices, Echelle spectrometer represents a very powerful tool for elemental analysis as demonstrated by Haisch et al.[25], who found substantial improvement in the detection limits for several elements with an Echelle system as compared with those obtained with a conventional Czerny- Turner system. A multi-spectrometer system developed by Ocean Optics is also being used for recording the broadband LIBS spectra for multielement analysis.

### 3.1.1 Echelle spectrometer

The principle of Echelle spectrometer has been described by Detalle et al[24]. It has focal length of 25 cm with a numerical aperture 1:10 and a quartz prism positioned in front of the grating separates the different orders of spectra and produces a two dimensional pattern. The flat image plane is $24.85 \times 24.85$ mm$^2$. This system provides maximum resolution in the wavelength range between 200 to 780 nm. The linear dispersion per pixel ranges from 0.005 nm (at 200 nm) to 0.019 nm (at 780 nm), which is based on the spectral resolution $\lambda/\Delta\lambda$ =40,000. The detector in this system is an ICCD camera, having a CCD array of $1024 \times 1024$ pixels ($24 \times 24$ $\mu$m$^2$) and a microchannel plate. A fast pulse generator delivers a 5 ns pulse to the intensifier to ensure synchronization of the measurements with the laser pulse. The spectral response in a particular order of diffraction of Echelle spectrometer is non-linear. It is measured using the blackbody radiation from a deuterium lamp, where maximum sensitivity is found in the center of the given order. Each diffraction order has similar shape but a different sensitivity, which requires a correction factor, when the measurement is made in different spectral range with different sensitivity. Normally, a blackbody radiation calibration spectrum is recorded to obtain the intrinsic response of the Echelle/ICCD system, which is then used to normalize the acquired spectrum.

### 3.1.2   Multi- spectrometer

A broadband multi- spectrometer suitable for recording LIBS has been developed by Ocean Optics Inc. Model LIBS 2000⁺. In this system a bundle of seven fibers (Diameter ~ 600 μm) collects the emission from the plasma spark. A lens is placed in front of the fiber bundle such that the plasma spark is sufficiently defocused for each fiber to collect the same emission and to eliminate any spatial effects. Each of the fibers is connected to an individual high resolution (0.1 nm) spectrometer covering different broadband spectrum from 200 – 980 nm spectral range (Ocean Optics Inc. LIBS 2000⁺). This becomes helpful in the multi element analysis of an inhomogeneous sample. Using either of the spectrometers LIBS spectra were collected with a 1.5 - 2 μs delay after plasma formation to eliminate plasma continuum effects. This system has been successfully utilized by De Lucia for detection of energetic materials[16].

## 4.   FIBER OPTIC LIBS PROBE

LIBS is most suitable for field based industrial applications, which include real time, on line analysis of material for process control and monitoring. Most of the experimental techniques discussed so far are laboratory based, where plasma is generated by focusing the high intensity laser beam on the sample surface with an assembly of lenses and the light emitted from the plasma is collected by either the same assembly of lenses or a separate assembly of lenses to be focused on the entrance slit of the spectrometer for further analysis[10]. Such an experimental set up is not well suited for field measurements, which require a flexible optical access to the test facility and minimal on-site alignment. Recent advances in fiber optic materials have opened up many new areas of applications for the LIBS technique. Using optical fiber, one can send the laser beam to the desired location and perform remote measurements. This system is most suitable for the study of solid, liquid and gaseous samples particularly hazardous chemical and biological samples not only in the laboratory but in the field also.

A schematic diagram of the fiber optic LIBS probe is shown in Fig.4[11]. The second harmonic (532 nm) of a pulsed Nd: YAG laser (Big Sky, Model CFR 400) operating at 10 Hz, with pulse duration 8 ns, beam diameter 7 mm, and the full angle divergence 1.0 m rad was directed into the optical fiber by a 532/1064-nm beam splitter and a 532 nm dichroic mirror. A specially coated 45° dichroic mirror (DM), which reflects at 532 nm and transmits in wavelength ranges 180-510nm and 550-1000 nm, was used to reflect the laser beam and to transmit the LIBS signal to the detection system. This simple design avoids any damage to the detector from the reflected laser light. To transmit sufficient laser energy below the damage threshold of fiber optic cable, the laser beam was focused 5 mm away from the fiber tip via a 10 cm focal length lens. A cap with a 0.8-mm pinhole was placed at the fiber input end to avoid the possibility of damage to the core and cladding of the fiber. The laser beam transmitted through the optical fiber was collimated with a 10 cm focal length lens and then focused on the sample with a 5-cm focal-length lens. The same lenses and optical fiber assembly were used to collect the emission from the laser-induced plasma and the collimated emission was passed through the dichroic mirror and focused onto an optical fiber bundle with a 20-cm focal length lens. The fiber bundle, a round-to-slit type, consists of 78 fibers, each having a 100 µm diameter and a 0.16 numerical aperture (NA). The slit-type end of the fiber bundle delivers the emission to the entrance slit of a 0.5-m focal length spectrometer (Model HR 460, JOBIN YVON-SPEX) equipped with a 2400 lines/mm grating blazed at 300 nm. An intensified charge couple device (Model ITF/CCD, Princeton Instruments) was used as the detector with its controller (Model ST 133, Princeton Instruments). A programmable pulse delay generator (MODEL PG-200, Princeton Instruments) was used to gate the ICCD and the data acquisition was under the control of a computer (Dell Dimension M 200a) running the WinSpec/32 software (Princeton Instruments). Multiple (100) lasers shot spectra were stored in one file, where fifty such spectra were recorded for analysis to get an average area/intensity value for the spectral lines under investigation.

The fiber optic probe discussed above is not truly field-portable LIBS instruments, due to the non- portability of the lasers employed and the length of optical fibers is also not unlimited (several meters in the above cases) due to signal and laser radiation attenuation in the optical fibers. A real field-portable LIBS instrument can only be realized by using a battery power supply, optical fibers and a miniature laser. Such a device was first developed in the research group of Cremers at Los Alamos National Laboratory[26]. It had a weight of 14.6 kg and a compact size of 46 x 33 x 24 cm$^3$ to fit into a small suitcase. The hand-held probe employed a compact, low cost, passively Q-switched Nd: YAG laser for making it portable. The laser had a low pulse energy (15-20 mJ pulse at 1064 nm, 4-8 ns duration) and repetition rate (< 1 Hz), but it had the ability to operate from 12 V D.C. batteries. A spark was produced on the sample by focusing the laser with a 50-mm focal length lens of 12 mm diameter. A fused silica fiber optic bundle of 2 m length was used to collect the emission from the laser induced plasma (LIP) and transmit it to a 1/8-m spectrograph. The end of the fiber optic bundle was positioned 5 cm from the LIP. It was not necessary to focus the LIP light onto the fiber with a lens due to the already sufficient emission collected in this configuration. The spectrally resolved emission was recorded with a compact CCD system. A compact computer was used for data processing and storage. The performance of the portable LIBS device was compared with that of a laboratory-based system using lead-containing paint samples and soil samples containing barium, beryllium, lead, and strontium. The results were identical in all aspects, indicating that downsizing the instrument did not affect its analytical performance. A more compact portable LIBS instrument has been built in Winefordner's research group at the University of Florida[27] using rechargeable batteries. This device is useful for field application, where regular power supplies are not available.

## 5.  ENHANCEMENT IN THE SENSITIVITY OF LIBS

It has been realized that poor detection limit is the most serious limitation of the LIBS technique in comparison to other analytical techniques. Several research groups have made modifications in the experimental setup and used different techniques to improve the LIBS detection limits. Magnetic confinement of laser-induced plasma has been used to enhance the sensitivity of LIBS by ~1.5-2 times, which resulted in decrease in the limit of detection by a factor of half.[14] Fig.-5 shows the LIBS spectrum of aluminum alloy in the absence and presence of magnetic field, which shows enhancement in the line emission from different element present in aluminum alloy. The combined use of a pair of laser pulses to ablate the material and further excite the resulting plasma to enhance the sensitivity of LIBS has been found most promising. This technique is known as Dual-Pulse LIBS, or Repetitive Spark Pair (RSP), or Double-Pulse Excitation. In some cases the dual pulse is delivered by a unique laser[28] while in other experiments use of two different lasers has been reported[29]. The second technique is more flexible with spatial arrangement of the two laser beams, their pulse energies and the time delay between two pulses. The use of a single laser, however, makes the system more compact and avoids the problems of alignment between the two laser pulses, ensuring better reproducibility.

St-Onge et al.[29] reported the study of some parameters affecting the performance of dual-pulse LIBS on metal samples in air. They found that the volume of the emitting plasma increases under the effect of the second laser pulse resulting in signal enhancement. This is due to more uniform absorption of the second laser pulse, whose energy is then distributed over a larger volume. St-Onge et al[30] also used a UV laser pulse to increase the ablation of the sample and an IR laser pulse to maximize the heating efficiency. In this configuration a significant signal enhancement was noted the extent of which varied depending on the ionization state and energy levels giving rise to the spectral lines of interest. A correlation has been established between the observed increase in intensity and the theoretical increase expected as a result of the higher plasma temperature generated by a combination of the UV-IR pulses. The enhancement in the signal was found greater than predicted by the increased temperature. This shows that an increase in plasma volume is also contributing the enhancement in intensity of emission. In contrast to the above-mentioned studies, where the plasma obtained by the first ablating pulse is reheated by a second pulse, Stratis et al[31] used a pre-pulse parallel to the sample surface and focused it to form air-plasma, followed by a second ablating pulse perpendicular to the sample surface and delayed in time by a few microseconds. In this case increase in intensity of the spectral line was correlated with an increased mass ablation. They simultaneously measured the time-resolved, spatially integrated emission intensity from two directions- perpendicular to the target surface; and parallel to the target surface- resulting in a slight difference, which indicates the importance of the collection geometry in the LIBS measurements. Rai et al.[32] used two lasers (Nd: YAG) for LIBS experiments and the spatially integrated emission in the direction opposite to the direction of the laser beams was collected (Fig.6). LIBS signal was enhanced by more than 6 times, when the time separation between two laser pulses was ~2-3 μs (Fig.-7). Smith et al.[33] used a different technique to improve detection limits by applying selective elemental excitation with a tunable diode laser to the LIP. Tunable diode laser induced atomic fluorescence was used for selective isotope detection of uranium containing samples. In order to detect the fluorescence signal, two techniques were employed: (i) the fast wavelength scanning of the diode laser during the lifetime of the plasma produced by each shot of the ablating laser and (ii) the time-integrated measurement with the diode laser wavelength fixed at the isotope line center. The optimal experimental conditions were found by means of a systematic scanning of the pressure of argon in the experimental chamber. The limit of detection in the optimal conditions was of the order of 0.6 ppm. Many other combinations of LIBS and laser-induced fluorescence (LIF) have been reported in recent years[34-35] to enhance the performance of LIBS as an extremely sensitive technique.

## 6.  APPLICATIONS OF LIBS

LIBS is a versatile technique for detection and identification of elements in a variety of samples that cannot be easily analyzed by other spectroscopic methods. Each one of these situations requires a modification of the standard LIBS instrumentation to give the best results. In the following sections we describe some of the unusual experimental arrangements.

### 6.1  Environmental Monitoring

#### 6.1.1  Off gas emission

The detection of hazardous and toxic trace elements in the off gas from waste processing system is very important for public health. LIBS has been used for in-situ off gas monitoring by focusing the laser beam in the gas stream through a window and collecting the optical emission through an optical fiber. Neuhauser et al.[36] have tested an on line lead (Pb) aerosol detection system with aerosol diameters ranging between 10 and 800 nm and a detection limit of 155 μg m$^{-3}$ has been achieved. LIBS has also been demonstrated as a process monitor and control tool for waste remediation[21]. The toxic metals from three plasma torch test facilities were monitored and it was found that LIBS can be integrated with a torch-control system to minimize toxic metal emission during plasma torch waste remediation. The possibility of using metal hydride to calibrate metals in off gas emission was also investigated[37] by using a static sample cell to perform LIBS measurements and the signal was found to be affected by gas composition, gas pressure and laser intensity.

The use of LIBS as continuous emission monitor (CEM) requires the quantitative trace level determination of the toxic metals. A system has been developed to monitor the concentration of selected toxic metals in near real time[22]. The concentrations of Be and Cr were measured at all the tested metal levels while that of Cd was measured during medium as well as high metal feed tests and the concentration of Pb was measured only at high concentrations. It was concluded that the LIBS system can be used as a CEM to monitor only the concentrations of Be, Cr and Cd but further improvements in the sensitivity of this system is required  for monitoring the Pb, Hg, As and Sb.

#### 6.1.2  Study of soil, concrete and paint

Detection of contaminated soil and concrete is an important area of environmental applications of LIBS. Yamamoto et al.[26] used a portable LIBS system to detect toxic metals in soil and detection limits of Ba, Be, Pb and Sr were found to be 265, 9.3, 298 and 42 ppm, respectively. Cremers et al.[38] detected Ba and Cr in soil using an optical fiber probe for remote operation with limit of detection of 26 ppm and 50 ppm for Ba and Cr, respectively. The effect of matrix was also studied in a soil sample. The limits of detection for Pb and Ba in a sand matrix were found as 17 and 76 ppm (by weight), respectively with a precision of 7% RSD whereas those in the soil were 112 and 63 ppm respectively with 10 % RSD. The LIBS signal was found to be affected by chemical speciation as well as matrix composition and its accuracy could be degraded if calibrations were not matrix specific.

Pakhomov et al.[39] have applied LIBS for the detection of Pb in contaminated concrete. A time resolved LIBS spectrum was recorded for the quantitative measurement of the Pb content in concrete. Pb calibrations were obtained by using the ratio of the integrated emission of lead line (405.78 nm) and that of an oxygen line (407.59 nm). It was found that the absolute Pb signal was independent of the laser pulse energy for laser energy between 250 and 400 mJ. The presence of Pb in the paint is a potential health threat, especially to children and Yamamoto et al.[26] have successfully demonstrated the feasibility of using LIBS to determine Pb in the painted surface.

### 6.1.3   Study of  radioactive elements

LIBS has also been used to monitor the level of radioactive elements in a process stream. Watcher and Cremers[40] found a detection limit of l00 ppm for uranium in solution. LIBS is preferable to other radiological measurements because nuclear detector may not be able to differentiate the radio nuclides U, Pu and Np. The LIBS spectra of U, Pu an Np were recorded in a globe box and the emission lines suitable for the detection of these radioactive elements were identified by Singh et al.[41]. The preliminary studies show that LIBS is suitable for the measurement of radioactive elements in waste stream. LIBS has also been used as a tool for detection of radiation embrittlement[9] in a nuclear power plant by determining the copper concentration in A533b steel. As copper is a key impurity contributing to radiation embrittlement, the Cu concentration in the steel may be an indicator of radiation embrittlement and expected material lifetime.

## 6.2 LIBS in Space Research

### 6.2.1 Rocket engine health monitor

Detection and characterization of metallic species in the exhaust plume of hydrocarbon-fueled rocket engines can indicate the onset of wear and / or corrosion of metal in the rocket engine. This information on engine wear obtained during engine operation is very useful, allowing the possibility of engine shutdown before any catastrophic failure. It has been observed that a catastrophic engine failure is generally preceded by a bright optical emission, which results from the erosion of metal from the engine parts. This is because of high temperature in the rocket plume (~2000 K), which partially vaporizes and atomizes the metal species, leading to atomic emission in the near ultraviolet and visible region (300- 760 nm). The performance of LIBS was evaluated by Rai et al.[42] in detecting the trace of elements in the fuel plume of a hybrid rocket engine simulator at Stennis Space Center, USA. Copper wire was inserted in the ignition chamber of engine and its vaporized trace was recorded in the rocket plume out side the exit nozzle (Fig.-8). The trace of copper was recorded near the nozzle exit during an initial fraction of a second, when the burnt-fuel plume started building up. However it decreased when plume attained its full length, high temperature and high speed. It was interesting to note that copper emission was observed throughout the plume away from the exit nozzle as well as in the luminous zone. This observation was attributed to better mixing of the metal vapor (away from the exit nozzle) along with decrease in background emission (due to luminous zone). It was found that the measurements made away from the luminous part of the plume could provide more meaningful information about the health of the rocket engine.

### 6.2.2 Probe for Mars expedition

Cremers et al[43]] are involved on evaluating the use of LIBS for future use on lander and rover to Mars. The main interest is the use of LIBS for stand off measurements of geological samples up to 20 meters from the instrument. The objective is to develop a very compact instrument operating at a remote distance from the target to detect at least 10 species in the rocks with detection limits < 100 pap including Ba, Li, Rb, and Sr with detection limit < 20 ppm. The ability to measure these separately in dust and pristine rocks is required. Minor and trace element composition are also important in determining the provenance of rocks and dusts. In an experiment under the simulated Martian atmospheric condition (5-12 mbar $CO_2$), it was noted that bulk matrix affected the calibration for Sr. Accuracy and precision were obtained in the detection of various other elements. Another project called MALIS (Mars elemental analysis by laser induced breakdown spectroscopy) is also in progress to demonstrate LIBS capability in Martian atmospheric condition[44].

## 6.3  Industrial Applications

Now a day all the metal producing industries are facing a major challenge of increasing productivity at reduced cost and maximizing the benefits from existing equipment. During refining, it is critical that operating parameters be adjusted and controlled so that the chemistry of the molten metal remains within predetermined limits. LIBS has been successfully used to get the composition of alloys (Fig.-5) along with their quantitative analysis in solid as well as in molten state[14, 45]. The analyte lines of Cu, Cr, Mn, Fe and Zn were used to obtain the calibration curve for their quantitative analysis. This type of analysis was performed by recording the spectra of alloy samples in the laboratory but LIBS has been found most suitable for field-based industrial applications, which include real time and online analysis of molten material for process control and monitoring.  Many groups[46-47] have used LIBS probe that uses one optical fiber for delivering the laser pulses to the target at a remote place for producing a micro-plasma as well as for collecting the resulting radiation from the LIP for quantitative elemental analysis.

## 6.4  Detection of Hazardous Chemical and Biological Samples

Among the different application of LIBS, the direct detection of hazardous chemical and biological samples such as chemical explosive and biological warfare agents in real time is a particularly challenging problem. Military has a vital interest in the development of field portable sensors to detect land mines, booby trap, remotely detonated munitions or unexploded ordnances. Similarly biological warfare involves the use of disease causing organism, toxins or other agents of biological origin to in capacitate, injure or kill human beings and animals as well as to destroy crops to weaken resistance to military attacks and to reduce the will to fight. Even the forensic laboratories have also interest in this technique for identification of culprit by detecting residue of gunshot on the hand and on cloths of the concerned person[18].

### 6.4.1 Detection of chemical explosive

F. De. Lucia et al[16] studied variety of energetic and explosive chemical materials using broad band LIBS system. First they studied the inorganic explosive material known as black powder, which has its typical component as charcoal, potassium nitrate, calcium sulfate, ammonium nitrate and sulfur. They also investigated several pure organic explosive materials such as RDX, HMX, PETN, NC and TNT.

They obtained the LIBS spectrum of black powder and found that spectrum has various peaks because of the presence of inorganic components, such as K with peaks at 691, 694,766 and 799 nm and Ca with peaks at 445, 558, 643 and 854 nm. The LIBS spectrum of organic explosive RDX has peaks due to C (247 nm), H (656 nm), N (746, 821 and 869 nm) and O (777 and 844 nm). All the other organic explosive materials also has similar peaks due to the presence of C, H, N and O. This shows that broadband spectrometer permits identification of all the atomic and molecular species over the spectral range 200 – 980 nm. However detection and identification of explosive chemical needs some other extensive development, which has two directions.

The first and most straightforward approach is through spectra matching based on a predetermined and assembled library of reference materials of interest. In this case one need the construction of a library of broad band spectra that would provide the basis for the comparison of the spectra for an unknown sample with those spectra contained in the library. This approach has been tested in the case of black powder successfully[16].

The second approach for the detection and identification of explosive is through use of the stoichiometery of the compound for discrimination by taking the ratio of the peaks of elements of interest. Anzano et al .[48] have used this technique successfully to short out the types of plastics. For example the intensity ratio of C line at 247 nm and the H line at 656 nm was found to be different for each plastic, whereas the LIBS spectra of all the plastics shared similar peaks, and the overall appearance of the LIBS spectra for the different plastics was highly similar. The peak intensity ratio of the elements was different to such an extent as to make differentiation of different plastics types possible. Similar approach has been used for identification of pure organic explosives[16]. Here one can get additional information from O and N peak intensity. However, care must be taken to avoid the effect of interference of atmospheric N and O. Another factor that affects this process of identification is change in plasma temperature, which finally reflected in change in the intensity ratio. F. C. De Lucia et al[16] have noted that LIBS plasma never starts ignition in the explosive during the experiment due to its short time duration of plasma. They have also used successfully the spectra peak intensity ratio as a basis for identifying compounds that have the same elements, but with different stoichiometries.

### 6.4.2 Detection of biological samples

The intensity ratio method used above has been found suitable in the identification of hazardous biological samples also. For this it is important to obtain atomic and ionic spectral lines that are both reproducible and specific for one sampled bacterial type. Aragon et al.[49] have found a lower dispersion in the line intensity ratio in comparison to single line intensity due to correction in statistical variation of the ablated mass and plasma characteristics. Generally accumulation of spectra induces a reduction in dispersion by averaging the shot-to-shot variations. Therefore the intensity ratio causes an improvement in measurements. Lines of interest must be chosen carefully. For example Carbon lines may be one of them for the bacteria due to its high density. The intensity ratio is indeed proportional to the ratio of Boltzmann factors and depends on the difference $\Delta E$ between the upper energy levels of the compounds. It has also been found according to Saha equation that lower relative standard deviation (RSD) values are obtained when the same kind of emission line (atomic or ionic) are used. Due to this reason S. Morel et al.[18] used cumulative intensity ratio (CIR) as one of important data. They used a series of 10 shots for different lines of two elements such as Mg and CN from the sample spectrum of Bacillus globigii BG-1. They obtained a plot for several cumulative ratio data versus the number of shots taken at the same as well as at different locations of the sample. It was found that ratio of cumulative measurements to number of shots fit a linear law with a good regression coefficient. The RSD was found relatively low 1.5 % for the ratio of Mg/CN. Table-1 shows the CIR of different elements having RSD data slightly higher for other elements, but below 10 %. The slope of this curve (CIR vs no. of shots) was found better for this analysis as they are different for different samples. Finally it has been reported that phosphorous (253.56 nm)/ carbon (247.85 nm) ratio is reproducible and is independent of any change in the parameter even during the preparation of two samples. CIR of P/C for different hazardous biological samples has been shown in table-2, which has clear-cut difference. This shows that CIR technique can be a useful method for identification of these samples using LIBS system.[18] Other methods of identification are also under investigation world wide.

## 6.5 Statistical Methods for Analysis of LIBS data

The success of LIBS as a field-portable detector for hazardous chemical and biological sample is dependent on the availability of some statistical method for rapidly analyzing the complex spectra obtained using the LIBS system. It needs development and the optimization of different statistical methods. Various statistical methods have been used to interpret spectroscopic data in many applications such as industrial process monitoring, study of forensic evidence and biomaterial identification[20, 50]. These analyzing techniques are known as chemo metrics. Some of the researchers have used these methods in the analysis of these hazardous materials and demonstrated their feasibility[20]. These statistical techniques include linear correlation, Principal component analysis (PCA) and soft independent modeling of class analogy (SIMCA). Munson et al.[20] have used these techniques to analyze several biological warfare agent simulants, biological inteferents (Pollens and molds) and chemical nerve agent simulants. They have used all the three statistical techniques. PCA is a data reduction technique that finds combinations of variables (Principal Components) that describe trends within the entire data set. From each PCA model is recovered the score and loadings of the individual principal components. Scores for each principal component describe the variation in the samples comprising the data set, while loadings describe the correlations among the variables. PCA data is typically depicted using a scores plot, a graph that relates the scores of the various components. In SIMCA, a PCA model is defined for each known sample category. The resulting model determines a residual variance for each category (tightness of the sample cluster) and object. The resulting values are used in a statistical F-test to determine the most probable category membership of unknown samples. The detailed discussion of the techniques of PCA and SIMCA can be obtained from the literature[51-52].

They have found that these analyzing techniques are partially successful in differentiating the samples using single shot emission spectra. Best results were attained using spectral averages and variance weighting methods. It was found that identification of these samples using LIBS spectra is a big challenge, because of similarities in its elemental composition. The complete discrimination of samples is possible only after developing a better statistical modeling and data pre-treatment procedures. It has also been realized that further investigation is needed to find the sources of shot to shot variations particularly plasma temperature, plasma volume and matrix effects.

## 7.   ANALYTICAL PERFORMANCE OF LIBS

The observed analytical figures of merit (precision, accuracy, and LOD) in LIBS experiments are inferior to those for other atomic spectrometric techniques such as ICP-AES and ICP-MS. The main reason for this deficiency is the multitude of experimental parameters that influence the analytical signal. These factors include laser wavelength, laser power, incidence angle, pulse-to-pulse variation, beam profile, beam shape, pulse duration, sample matrix, freshness of surfaces for solid samples, purging gas, ambient pressure etc. Even the experimental arrangement and sampling geometry affects the LIBS measurements significantly[53]. These factors have to be identified and optimized for better analytical performance. Spatial and temporal dependence of emission signals from the laser- induced plasma (LIP) have been studied by several research groups[54-55]. It has been realized that the investigation of methods for measuring relative mass removal in the laser-induced plasma would probably continue for a long time and will be valuable, not only for LIBS experiments, but also for LIP-ICP-AES and LIP-ICP-MS experiments. It is expected that the fundamental studies of ablation and the excitation processes in the laser-induced plasma using different wavelengths and ultra-short pulses would enhance the analytical capabilities of LIBS. A good review on these aspects has been published by Rusak et al.[7].

The major limitations of LIBS for practical applications result from self-absorption, line broadening, and the high intensity of the background continuum along with strong matrix effects [68]. Some of these limitations can be minimized or avoided by working in a controlled atmosphere and using time-resolved spectroscopic measurements or time-integrated and spatially resolved measurement techniques. In time-resolved spectroscopy, the temporal evolution of the plasma is obtained by recording the plasma emission spectra at various delay times. The LIBS spectra of magnesium in liquid matrix were reported by Rai et al.[56] who found that 500 ns after the laser irradiation, the observed spectrum consisted of continuum and ion emission lines, as the plasma temperature was high, but after 10 μs the intensity of the continuum and ion lines decreased, and that of lines due to neutral atoms increased as a result of electron ion recombination.

The purpose of time-integrated and spatially resolved spectroscopy is to measure the emission from the LIP at different positions of the plasma. Lee et al.[12-13, 57] have carried out experiments on copper and lead using ArF, XeCI and Nd: YAG lasers. They found that the plasma consisted of two distinct regions when the ambient pressure was reduced below 50 torr in air or argon atmosphere. The region near the target surface referred to as the inner sphere plasma, emitted a strong signal of copper ions and continuum background. The other region referred to as the outer sphere plasma, surrounded the inner sphere plasma and emitted blue-green copper atomic lines with a relatively low background continuum and without ion lines. They reported the LOD values in the range of several parts per million to hundreds of parts per billion in solid samples, while the R.S.D. values varied from a few percent to 80%. Wachter and Cremers[40] have examined the effects of the laser pulse repetition rate, the detector gating, and the number of averaged laser shots on the precision. It was found that the precision increased with repetition rate and total number of laser pulses averaged, but was independent of the gating parameters. It was also found that the R.S.D. decreased from 13.3% for 50 laser shots to 1.8% for 1600 laser shots. The freshness of sample surface also affects precision and the lowest R.S.D. has been obtained, when each shot samples a totally new portion of the material[58]. Eppler et al.[59] found an increase in the precision using a cylindrical lens instead of a spherical lens, but the R.S.D. was independent of the choice of lens. The reduction of the R.S.D. was attributed to the greater amount of material sampled by the cylindrical lens. Castle et al.[27] have systematically studied variables that influence the precision of LIBS measurements with special emphasis on the effect of temporal development of the emission, the sample translational velocity, and the number of spectra accumulated, laser pulse stability, detector gate delay, surface roughness, and the use of background correction.

## 8. CONCLUSION

Developments of LIBS techniques have been very rapid in the recent years and commercial instruments are coming up for application in process monitoring in many industries as well as in various field applications. The prototypes of miniaturized version of LIBS have already been demonstrated and it is hoped that these will be commercially available in near future. However in spite of many advantages LIBS techniques lag behind the conventional analytical techniques in the terms of sensitivity but work is in progress to circumvent this shortcoming by a careful assessment of the experimental parameters that influence the LIBS signal. The miniaturized version of LIBS will be very much applicable for detection and identification of hazardous chemical and biological samples, in bio-medical applications as well as in forensic science. Development and optimization of data analysis techniques is also required to make the system more accurate in identification of complex organic molecules having similar elemental composition such as explosive and biological samples. Finally it seems that applicability of LIBS will increase many fold after the development of a miniaturized LIBS system with enhanced sensitivity and relevant statistical methods for data analysis.

**Table-1**    Reproducibility of CIR on the same pellet but with different compounds. (Ref. # 18 )

| S. No. | CIR | Slope (RSD) | RSD for shot No.   10 (%) | RSD after 10 shots (%) |
|--------|-----|-------------|---------------------------|------------------------|
| 1 | Mg/CN | 0.64(1.5%) | 3.46 | 1.85 |
| 2 | Ca/CN | 2.49(8.45%) | 9.65 | 8.46 |
| 3 | K/CN | 0.065(6.81%) | 9.02 | 6.84 |
| 4 | Ca/K | 38.05(2.44%) | 8.76 | 1.87 |
| 5 | Ca/Mg | 3.92(7.81%) | 8.75 | 7.51 |
| 6 | Mg/K | 9.76(6.78%) | 5.54 | 6.15 |

**Table-2**    Biological agents and related pathogenic agents with their CIR number obtained from the ratio of phosphorus and carbon [P (253.56 nm)/C (247.85 nm)](Ref. # 18)

| S. No. | Biological Agents | Related Pathogenic Agents | CIR  (P/C) Approx. |
|--------|-------------------|---------------------------|--------------------|
| 1 | Bacillus globigii (BG-1) | Bacillus anthracis | 0.98 |
| 2 | Bacillus globigii (BG-2) | Bacillus anthracis | 1.10 |
| 3 | Bacillus thurengensis | Bacillus anthracis | 0.50 |
| 4 | Escherichia coli | Yersinia pestis | 0.80 |
| 5 | Staphylococcus aureus | Staphalococcus epidermis | 1.01 |
| 6 | Proteus mirabilis | Proteus mirabilis (Causes urinary and gastro intestinal infections) | 0.62 |
| 7 | Popplar pollen | | 0.10 |

**REFERENCES**

1.      F. Brech and L. Cross, Appl. Spectrosc. **16** 59 (1962).

2.      E. R. Runge, R.W. Minck and F.R. Bryan, Spectrochim. Acta. **20B** 733 (1964).

3.      L. J. Radziemski and D. A. Cremers, "Laser-Induced Plasma and Applications", Marcel Dekker, New York, (1989) pp 295-325.

4.      L. Moenke-Blankenburg, "Laser Microanalysis", (Eds.) J. D. Winefordner, I. M. Kolthoff , John Wiley & Sons, New York, (1989).

5.      L. Moenke-Blankenburg, "Laser in Analytical Atomic Spectroscopy", (Eds.) J. Sneddon, T. L Thiem, Y. I. Lee, Wiley-VCH, New York, (1997) pp. 125-195.

6.      S. A. Drake and J. F. Tyson, J. Anal. At. Spectrom. **8** 145 (1993).

7.      D.A.Rusak, B.C.Castle, B.W.Smith and J.D.Winefordner, Crit. Rev. Anal. Chem. **27** 257 (1997).

8.      K. Song, Y. I. Lee and J. Sneddon, Appl. Spectrosc. **32** 183 (1997).

9.      F. Y. Yueh, J. P. Singh and H. Zhang, "Encyclopedia of Analytical Chemistry" (Ed.) R. A. Meyers, Vol. 3, Wiley, New York, (2000) p 2065

10.     O. Samek, D. C. S. Beddows, J. Kaiser, S. V. Kukhlevsky, M. Liska, H. H. Telle and J. Young, Opt. Eng. **38** 2248 (2000).

11.     A. K. Rai, V. N. Rai, F. Y. Yueh and J. P. Singh, "Trends in Applied Spectroscopy" Vol. **4,** Research Trends, Trivandrum, India, (2002) p 165.

12.     Y. I. Lee, K. Song and J. Sneddon "Laser in Analytical Atomic Spectroscopy", (Eds.) J. Sneddon, T. L. Thiem and Y. I. Lee, Wiley-VCH, New York, (1977) p. 197

13.     Y. I. Lee, Y. J. Yoo and J. Sneddon, Spectroscopy **13** 14 (1998).

14.     V. N. Rai, A. K. Rai, F. Y. Yueh and J. P. Singh, Appl. Opt. **42** 2085 (2003).

15.      F. Y. Yueh, V. N. Rai, J. P. Singh and H. Zhang, Paper # AIAA-2001-2933, 32[nd] AIAA Plasmadynamics and Laser Conference, 11-14 June (2001), Anaheim, CA, USA.

16.     F. C. De Lucia, Jr. R. S. Harmon, K. L. Mc Nesby, R. J. Winkel Jr. and A. W. Miziolek, Appl. Opt. **42** 6148 (2003).

17.     J. D. Hybl, G. A. Lithgow and S. G. Buckley, Appl. Spectrosc. **10** 1207 (2003).

18.     S. Morel, N. Leone, P. Adam and J. Amouroux, Appl. Opt. **42** 6184 (2003).

19. C. R. Dockery and S. R. Goode, Appl. Opt. **42** 6153 (2003).

20. C. A. Munson, F. C. De Lucia Jr., T. Piehler, K. L. Mc Nesby and A. W. Miziolek, Spectrochimica Acta **60B** 1217 (2005).

21. J. P. Singh, F. Y. Yueh, H. Zhang and R. L. Cook, Process Controll and Quality **10** 247 (1997)**.**

22. H. Zhang, F. Y. Yueh and J. P. Singh, Appl. Opt. **38** 1459 (1999).

23. H. E. Bauer, F. Leis and K. Niemax, Spectrochim. Acta. **53B** 1815 (1998).

24. V. Detalle, R. Heon, M. Sabsabi and  L. St. Onge, Spectrochim. Acta **56B** 1011 (2001).

25. C. Haisch, U. Panne and R. Niessner, Spectrochim Acta **53B** 1657 (1998).

26. K. Y. Yamamoto, D. A. Cremers, M. J. Ferris and L. E. Foster, Appl. Spectrosc. **50** 222 (1996).

27. B. C. Castle, A. K. Knight, K.Visser, B.W.  Smith and J. D. Winefordner, J. Anal. At. Spectrom. **13** 589 (1998).

28. A. E. Pichahchy, D. A. Cremers and M. J. Ferris, Spectrochim. Acta **52B** 25 (1997).

29. L. St-Onge, M. Sabsabi and P. Cielo, Spectrochim. Acta **53B** 407 (1998).

30. L. St-Onge, V. Detalle and M. Sabsabi, Spectrochim Acta **57B** 121 (2002).

31. D. N. Straits, E. L. Eland and S. M. Angel, Appl. Spectrosc. **54** 1270 (2000).

32. V. N. Rai, F. Y. Yueh and J. P. Singh, Appl. Opt. **42** 2094 (2003).

33. B. W. Smith, A. Quinter, M. Bolshov, K. Niemax, Spectrochim. Acta **54B** 943 (1999).

34. I. B. Garnishing, S. A. Baker, B. W. Smith and J. D. Winefordner, Spectrochim Acta **52B** 1653 (1997).

35. H. H. Telle, D. C. S. Beddows, G. W. Morris and O. Samek, Spectrochim. Acta. **56B** 947 (2001).

36. R. E. Neuhauser, U. Panne, R. Neissner, G. A. Petrucci, P. Cavalli and  N. Omenetto, Annal. Chim. Acta. **346** 37 (1997).

37. J. P. Singh, H. Zhang, F. Y. Yueh and K. P. Karney, Appl. Spectrosc. **12** 764 (1996).

38. D. A. Cremers, J. E. BarefieldII and A. C. Koskelo, Appl. Spectrosc. **49** 857 (1995).

39. A. V. Pakhomov, W. Nichols and J. Borysow, Appl. Spectrosc. **50** 880 (1996).

40. J. R. Watcher and D. A. Cremers, Appl. Spectrosc. **41** 1042 (1987).

41.     J. P. Singh, F. Y.Yueh, H. Zhang and K P. Karney, Rec. Res. Dev. Appl. Spectrosc. **2** 59 (1999).

42.     V. N. Rai, J. P. Singh, C. Winstead, F.Y.Yueh and R. L. Cook, AIAA Journal **41** 2192 (2003).

43.     D. A. Cremers, R. C. Wiens, M. J. Ferris, R. Brennetot and S. Maurice, Trends in Optics and Photonics Vol. **81** Laser-Induced Plasma Spectroscopy and Applications, OSA Technical Digest (2002) p. 5.

44.     R. Brennetot, J. L. Lacour, E. Vors, P. Fichet, D. Vailhen, S. Maurice and A. Rivoallan, Trends in Optics and Photonics, Vol. **81**, Laser induced plasma Spectroscopy and Applications, OSA Technical Digest (2002) P9.

45.     A. K. Rai, F. Y. Yueh and J. P. Singh, Rev. Sci. Instrum. **73** 3589 (2002).

46.     C. Aragon, J. A. Aguilera, J. Campos, Appl. Spectrosc. **47** 606 (1993).

47.     L. Peter, V. Sturm and R. Noll, Appl. Opt. **42** 6199 (2003).

48.     J. M. Anzano, I. B. Gormushkin, B. W. Smith and J. D. Winefordner, Polym. Eng. Sci. **40** 2423 (2000).

49.     C. Aragon, J. A. Aguilera and F. Penalba, Appl. Spectrosc. **53** 1259 (1999).

50.     N. S. Foster, S. E. Thompson, N. B. Valentine, J. E. Amonette and T. J. Jhonson, Appl. Spectrosc. **58** 203 (2004).

51.     M. A. Sharaf, D. L. Illman, B. R. Kowalski, P. J. Elving, J. D. Winefordner (Ed.) Chemometrics, Chemical Analysis, Vol. **82**, John Wiley & Sons New York 1986.

52.     G. H. Dunteman, Principal Components Analysis, Quantitative Applications in the Social Sciences, Vol. 69 SAGE Publications Inc. Newbury Park CA 1989.

53.     R. A. Multari, L. E. Foster, D. A. Cremers and M. J. Ferris, Appl. Spectrosc. **50** 1483 (1996)

54.     B. C. Castle, K. Visser, B. W. smith and J. D. Winefordner, Appl. Spectrosc. **51** 1017 (1997).

55.     K. Song, H. Cha, J. Lee and Y. I. Lee, Microchem. J., **63** 53 (1999).

56.     V. N. Rai, H. Zhang, F. Y. Yueh, J. P. singh and A. Kumar, Appl. Opt. **42** 3662 (2003).

57.     Y. I. Lee, T. L. Thiem, G. H. Kim, Y. Y. Teng and J. Sneddon, Appl. Spectrosc. **46** 1597 (1992).

58.    R. Wisbrun, I Schechter, R. Niessner, h. Schroder and K. L. Kompa, Annal. Chem. **66** 2964 (1994).

59.    A. S. Eppler, D. A. Cremers, D. D. Hickmott, M. J. Ferris and A. C. Koskelo, Appl. Spectrosc. **50** 1175 (1996).

**FIGURE CAPTIONS**

1.    Schematic diagram of experimental system for recording LIBS of solid samples (Ref. # 14   ).

2.    Schematic diagram of experimental system for recording LIBS of liquid samples. In these experiments plasma is produced on the surface of liquid or liquid jet  (Ref. # 11 & 15 ).

3.    LIBS calibration system for gaseous samples (a) Open system (b) Closed system. Rest of the instrument is similar as shown in fig. 1 & 2  (Ref. # 21 & 22)

4.    Schematic diagram of fiber optic probe for remote analysis of LIBS signal from a sample. Similar system is used for recording LIBS from samples submerged under water or molten liquid samples in industries.  In the molten metal case optical fiber is passed through a ceramic tube and dry air, $N_2$ or argon gas is passed through it for creating a bubble at the place of plasma formation. In some cases fiber is kept close enough to the sample for creating plasma without a focusing lens (Ref. # 11).

5.    LIBS spectra of Al alloy recorded at 25 mJ laser energy pulse at a gate delay of 5 μs and a gate width of 2 μs: (a) no magnetic field, (b) magnetic field present. (Ref. # 14)

6.    Schematic diagram of optical system for recording the LIBS spectrum of liquid sample (Jet) in double laser pulse excitation mode. Rest of the system such as dispersing device, detectors and data acquisition components are same as shown in Fig.-1 (Ref. # 32).

7.    Emission spectra of Mg (5 ppm) in double pulse excitation mode with a change in interpulse delay (0 and 3 μs) between lasers (Laser1, 100 mJ; Laser2, 120 mJ; gate delay/gate width 10 μs/10 μs) (Ref. # 32)

8.    LIBS spectrum of rocket engine plume (Laser energy ~ 280 mJ and Gate delay / gate width 30/10 μs) with copper seed (Copper wire in ignition chamber) (Ref. # 42).

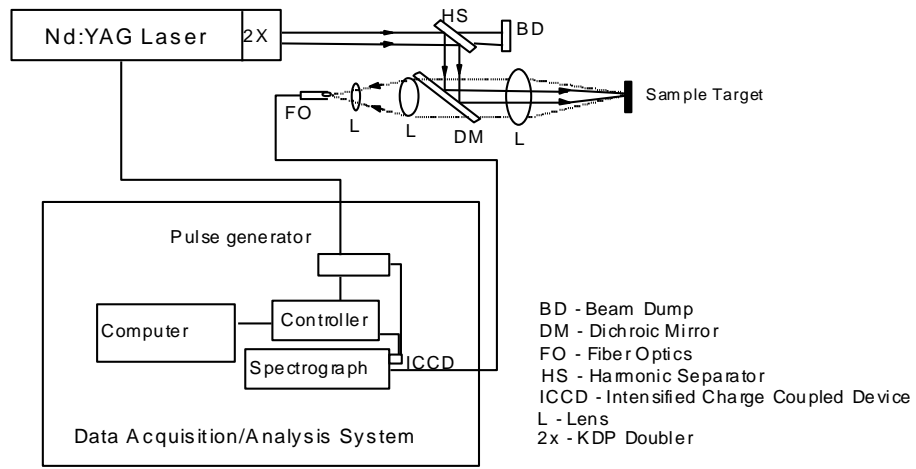

Fig.-1

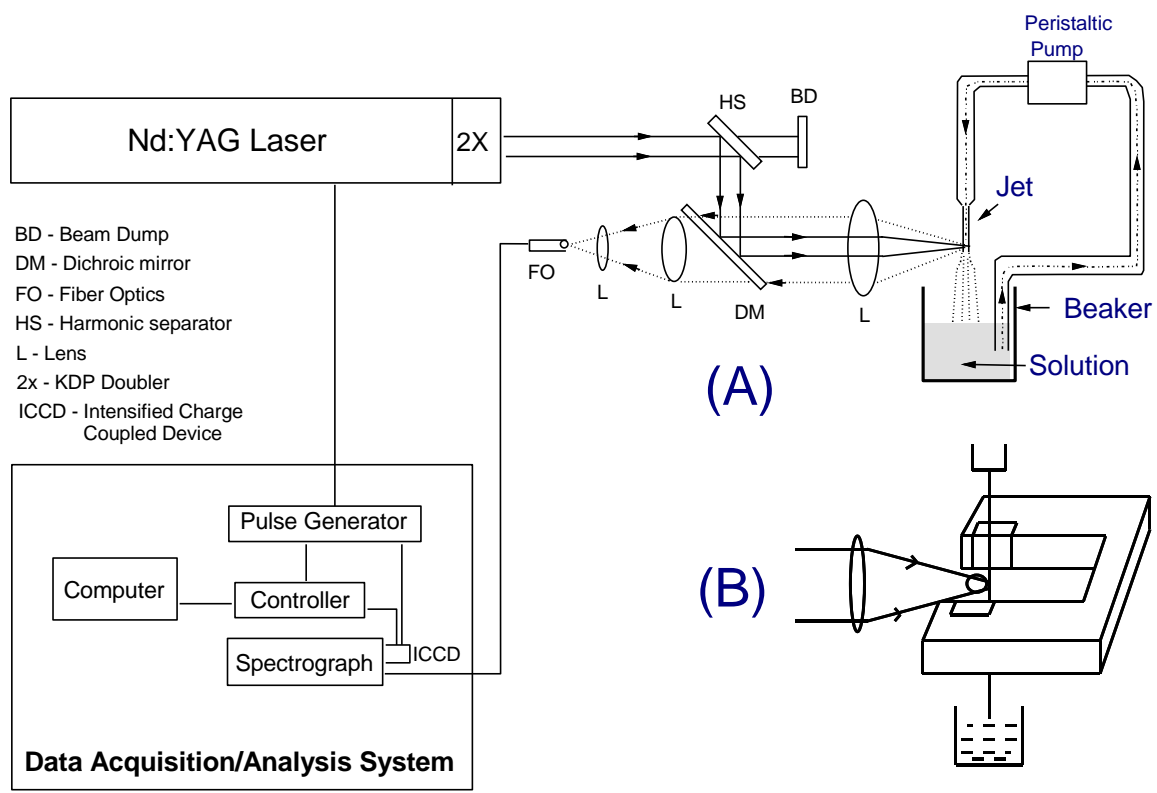

BD - Beam Dump
DM - Dichroic mirror
FO - Fiber Optics
HS - Harmonic separator
L - Lens
2x - KDP Doubler
ICCD - Intensified Charge
        Coupled Device

Fig.-2

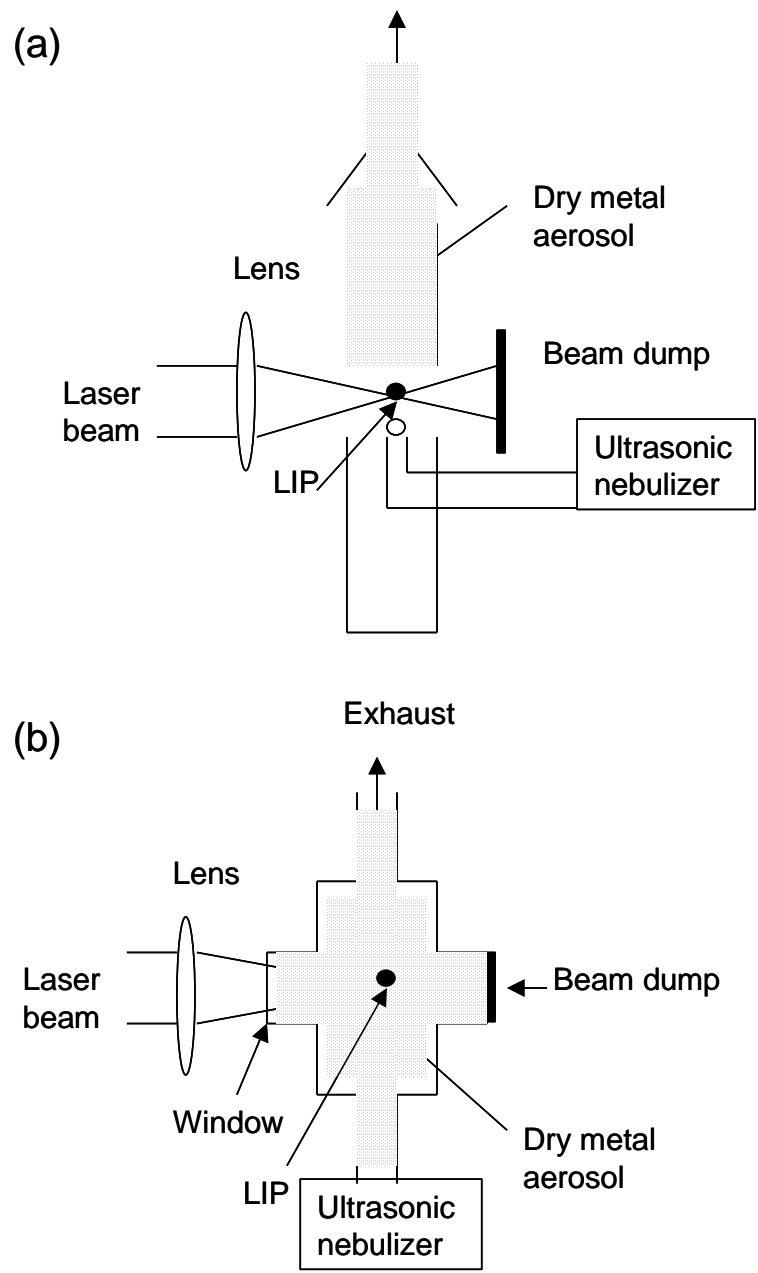

(a)

Dry metal aerosol

Lens

Beam dump

Laser beam

LIP

Ultrasonic nebulizer

(b)

Exhaust

Lens

Laser beam

Beam dump

Window

Dry metal aerosol

LIP

Ultrasonic nebulizer

Fig.-3

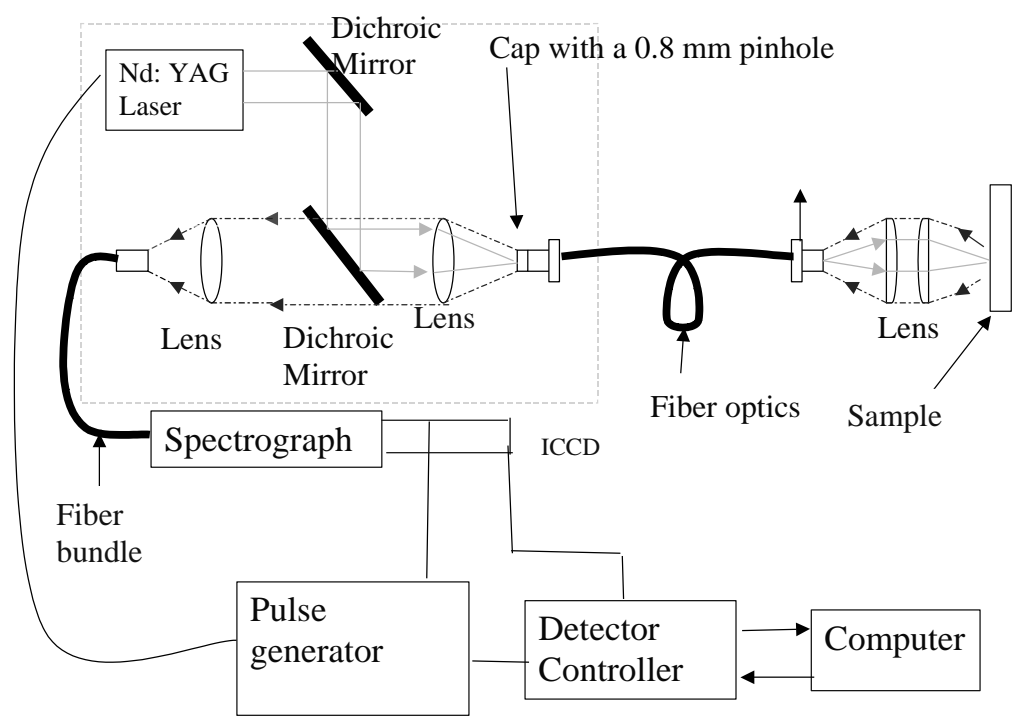

Fig.-4

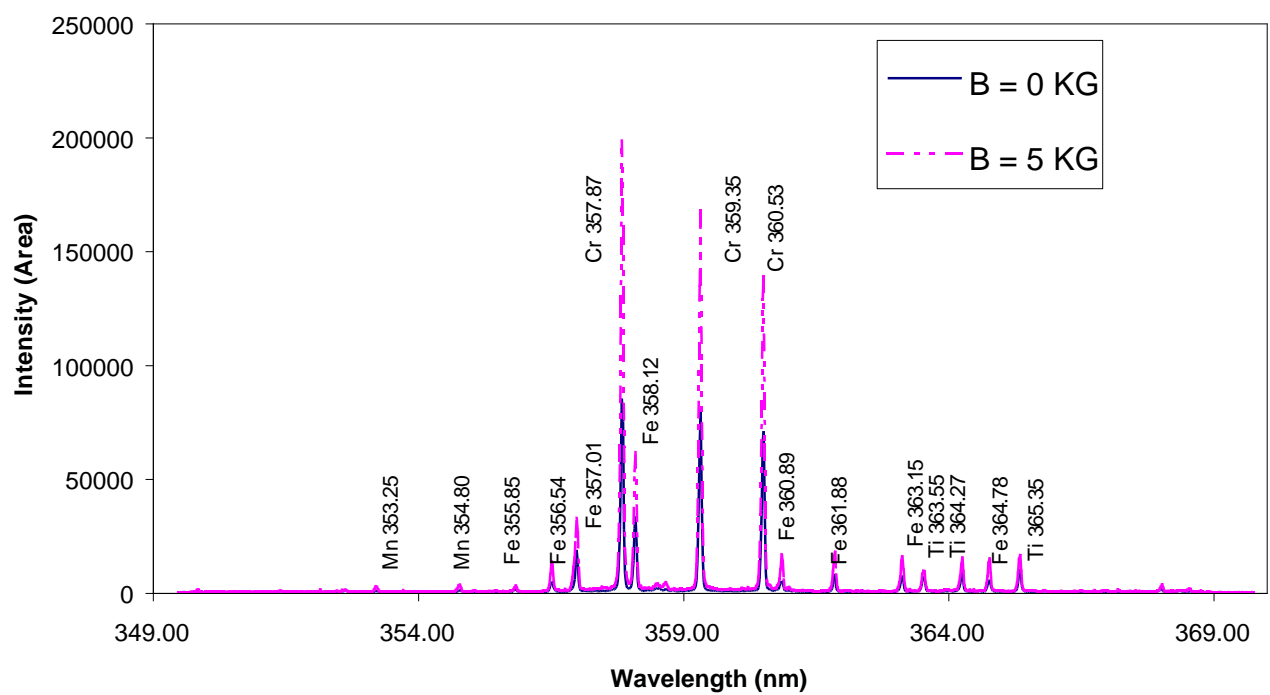

Fig.-5

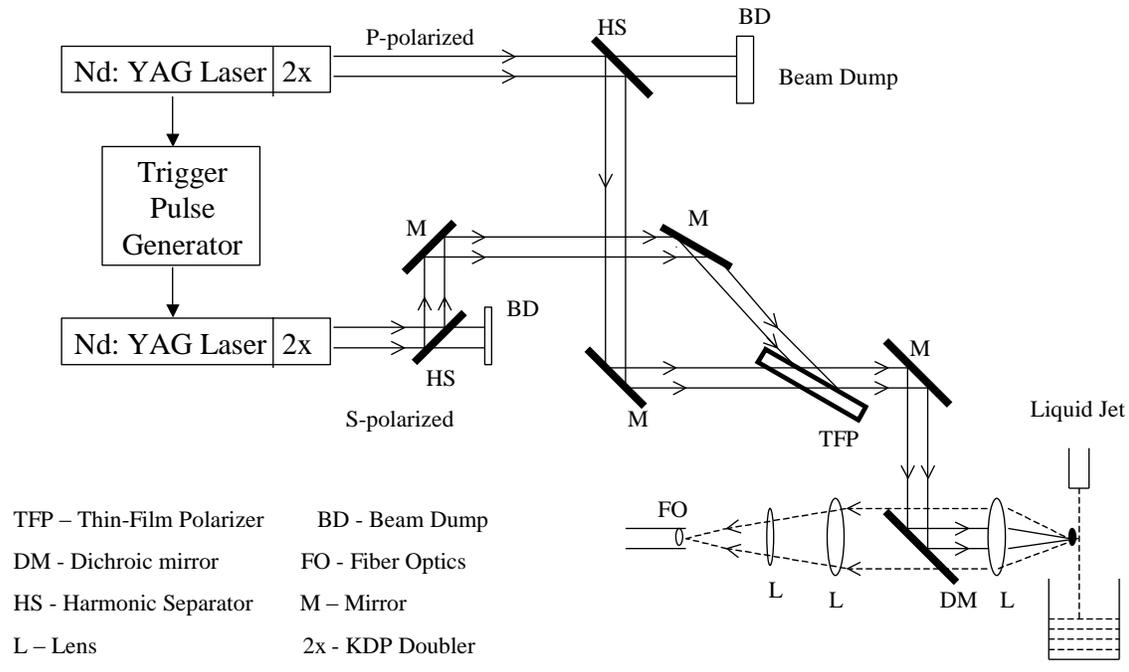

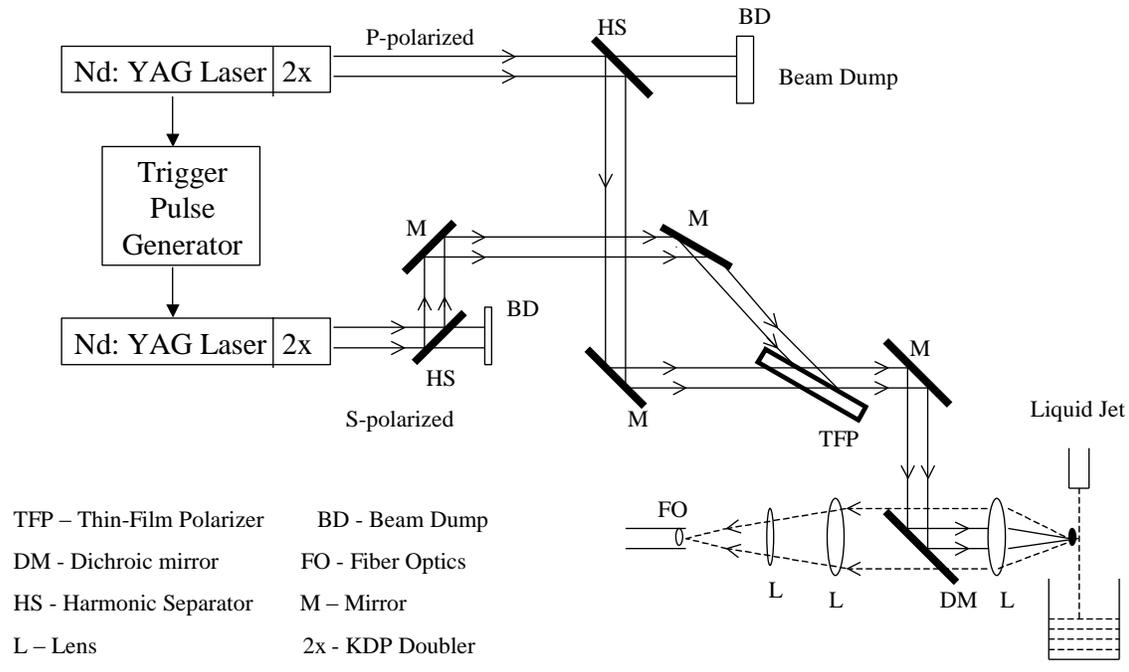

Fig.-6

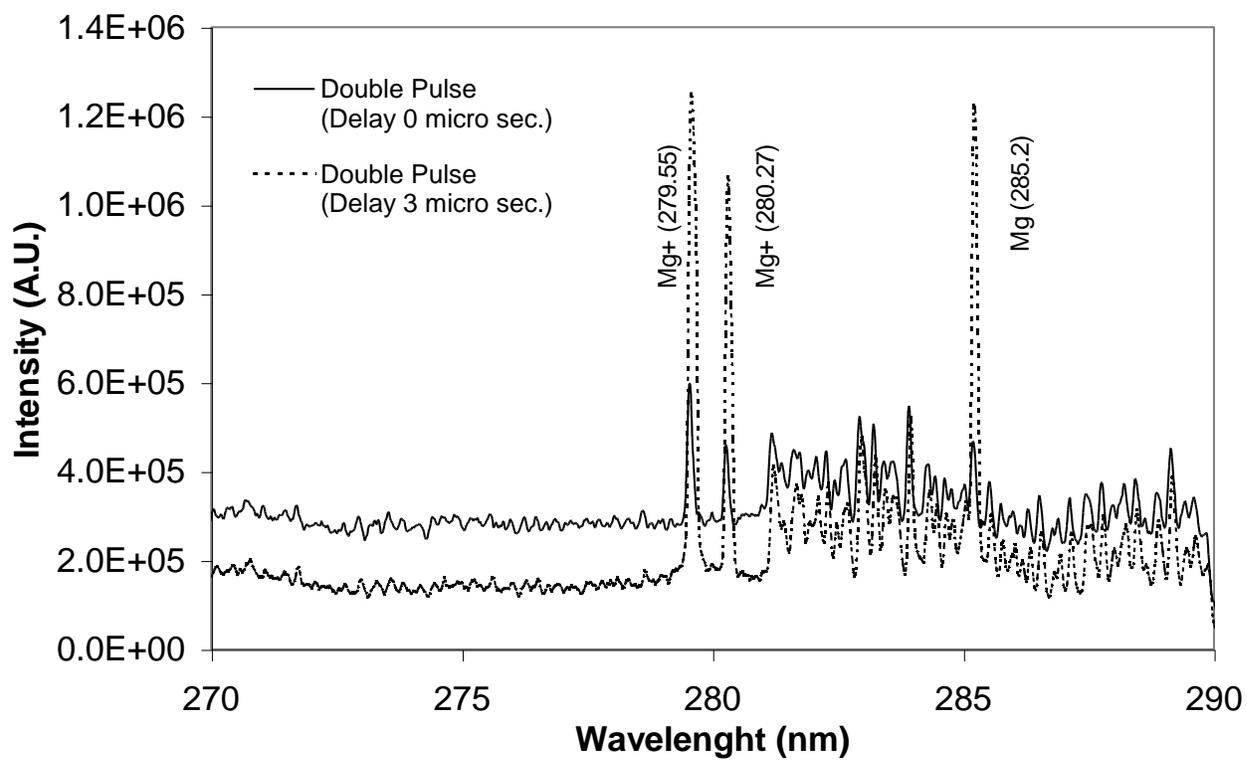

Fig.-7

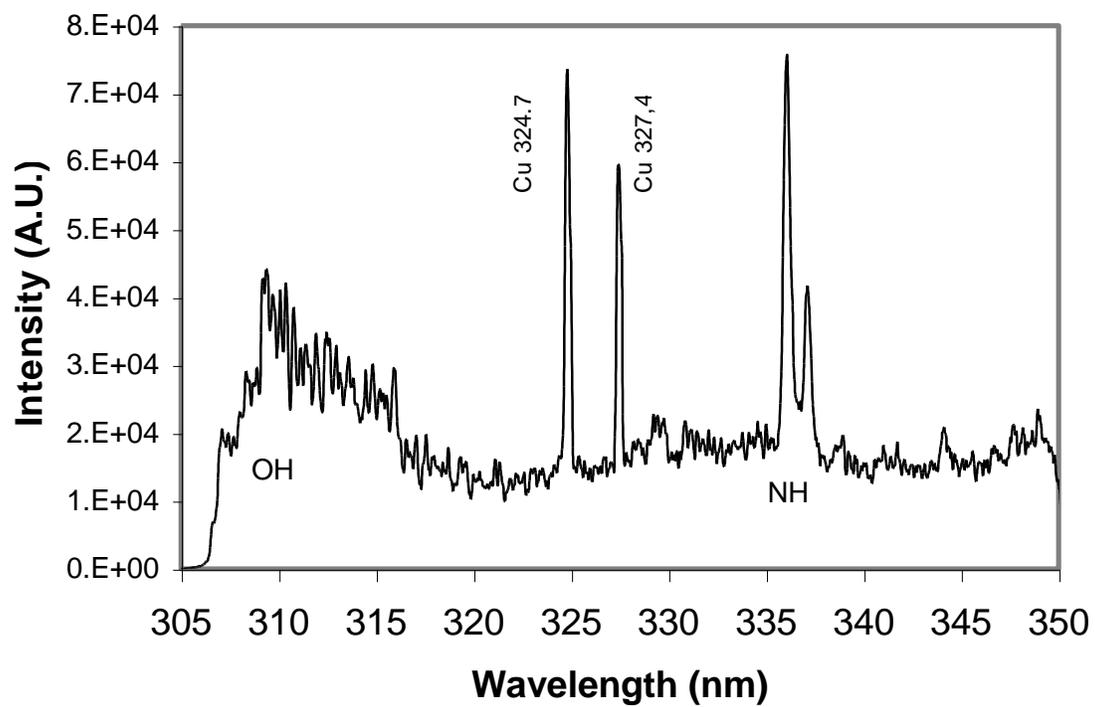

Fig.-8